\documentclass{emulateapj}
\usepackage{epsf}
\usepackage{times}

\shorttitle{SNe Ia IN A HIERARCHICAL GALAXY FORMATION MODEL}
\shortauthors{NAGASHIMA \& OKAMOTO}

\begin{document}

\title{Type Ia supernovae in a hierarchical galaxy formation model: the Milky Way
}

\author{Masahiro Nagashima\altaffilmark{1,2}
and Takashi Okamoto\altaffilmark{1,3}}
\email{masa@scphys.kyoto-u.ac.jp}
\altaffiltext{1}{Department of Physics, University of Durham, South Road, Durham DH1 3LE,
United Kingdom}
\altaffiltext{2}{Department of Physics, Graduate School of Science, Kyoto
University, Sakyo-ku, Kyoto 606-8502, Japan}
\altaffiltext{3}{
Division of Theoretical Astrophysics, National Astronomical
Observatory, National Institute of Natural Science, Mitaka, Tokyo 181-8588, Japan}

\begin{abstract} 
 We investigate chemical evolution in Milky Way-like galaxies based on
 the cold dark matter model in which cosmic structures form via
 hierarchical merging.  We introduce chemical enrichment due to type Ia
 supernovae (SNe Ia) into the Mitaka semi-analytic galaxy formation
 model developed by Nagashima \& Yoshii.  For the first time we derive
 distributions of stellar metallicities and their ratios in Milky
 Way-like galaxies treating chemical enrichment due to SNe Ia in a
 hierarchical galaxy formation model self-consistently.  As a first
 attempt, we assume all SNe Ia to have the same lifetime, and assume
 instantaneous recycling for type II supernovae (SNe II).  We find that
 our model reproduces well the metal abundance ratio [O/Fe] against
 [Fe/H] and the { iron metallicity distribution function} in the
 solar neighborhood.  This means that the so-called G-dwarf problem is
 resolved by the hierarchical formation of galaxies, and a gas infall
 term introduced in traditional monolithic collapse models to solve this
 problem is well explained by the mixture of some physical processes
 such as hierarchical merging of dark halos, gas cooling, energy
 feedback and injection of gas and metals into hot gas due to
 supernovae.  Our model predicts more oxygen-enhanced stars in bulges at
 [Fe/H] $\simeq 0$ than in disks.  This trend seems to be supported by
 recent observations while they have still uncertainties.  More data in
 number and accuracy will provide independent and important constraints
 on galaxy formation.  For the better understanding of the chemical
 enrichment due to SNe Ia in hierarchical galaxy formation, we discuss
 how physical processes affect the metal abundance ratio by varying the
 lifetime of SNe Ia, the star formation timescale and the strength of
 supernova feedback.  We find that the supernova feedback plays a key
 role among them and therefore there is no one-to-one correspondence
 between star formation histories and stellar metallicity-ratio
 distributions.
 \end{abstract}


\section{INTRODUCTION}   
The chemical compositions of stars and gas, as is widely known, provide
important clues to understanding the formation and evolution of
galaxies.  In particular the abundance ratio of $\alpha$-elements such
as oxygen to iron is a useful probe of the formation history of
galaxies.  The main sources of metals are considered to be supernovae
(SNe), especially type II SNe (SNe II) and type Ia SNe (SNe Ia).  They
have different abundance patterns, the former produces $\alpha$-elements
as well as iron and the latter produces hardly any $\alpha$-elements.
Therefore, the measurement of [$\alpha$/Fe] is closely related to the
ratio of SNe II to SNe Ia.  While the progenitors of SNe II are massive
stars larger than $\sim 10M_{\odot}$, those of SNe Ia are considered to
be white dwarfs in binary systems.  Therefore the explosion rate of SNe
II is almost proportional to the star formation rate of their progenitor
stars at that time, but there is a time lag between SNe Ia explosions
and the formation of their progenitor stars.  Since the evolution
timescales of those elements are independent and different, we can
obtain useful information on galaxy formation and star formation
histories from the abundance ratio of those elements.

Because SNe II explode almost instantaneously, the chemical evolution
process has usually been formulated by assuming that massive stars
immediately explode right after their formation (the instantaneous
recycling [IR] approximation)\citep{t80, lf83, lf85}.  In those works,
galaxies are usually assumed to grow up in monolithic clouds, in which
timescales of gas infall and star formation (SF) are typically treated
as free parameters.  Since it is possible to observe and resolve
individual stars in the solar-neighborhood, many studies have focused
on evolution of the galactic disk.  By adjusting those parameters and by
incorporating stellar population synthesis techniques \citep{ay86,
ay87}, detailed galaxy evolution models that reproduce observations have
been developed \citep{ayt92}.

In the framework of such traditional galaxy evolution models, several
authors explicitly considered chemical enrichment due to SNe Ia
\citep{gr83, mg86, mf89, tnyhyt95, pt95, ytn96}.  They included delayed
production of metals due to SNe Ia in the galaxy evolution models.  By
comparing their models with solar neighborhood stars in { the
metallicity distribution function (MDF)} and abundance ratio such as
[O/Fe] against metallicity, \citet{pt95} and \citet{ytn96} inferred a
typical lifetime of SNe Ia of 1.3$\sim$1.5 Gyr.  Although
\citet{ktnhk98} has claimed that SNe Ia start to explode at the age of
about 0.6 Gyr, \citet{gm04} recently derived that the lifetime should be
longer than about 1 Gyr at the 99 per cent confidence level by comparing
the cosmic star formation history with the cosmic SNe Ia rate.

Considered from the viewpoint of cosmological structure formation,
however, these models are still only phenomenological.  Recent
theoretical and observational studies of structure formation have
revealed that the universe is dominated in mass by cold dark matter
(CDM) and that baryonic objects such as galaxies form in virialized
objects of dark matter called dark halos.  In the CDM model, since
smaller-scale density fluctuations in the early universe have larger
fluctuation amplitudes, larger halos form through mergers of smaller
dark halos (the hierarchical clustering scenario).  { This suggests
that large galactic gas clouds would have formed at low redshift by
mergers of subgalactic clumps which have formed at higher redshift.
Therefore} we may have to modify our picture of chemical evolution in
galaxies as well as that of galaxies themselves.

Recently, based on a semi-analytic (SA) approach, galaxy formation
models in the hierarchical clustering scenario have been developed, in
which the formation histories of individual dark halos are followed by
using a Monte Carlo method and physical processes such as gas cooling
and star formation are taken into account in the histories of dark
halos.  This is a natural consequence of adopting the CDM model because
we are not able to freely assume formation histories of dark halos as
far as we would like to construct galaxy formation models consistent
with the CDM model.  Many authors have found that such SA models
reproduce well various characteristics of galaxies at the present and at
high redshift such as luminosity functions, gas fractions, size
distributions, and faint galaxy number counts \citep[e.g.,][]{kwg93,
cafnz94, clbf00, ngs99, ntgy01, nytg02, ny04, sp99, spf01}.  In recent
SA models, chemical enrichment is considered, but generally only with
the IR approximation.  \citet{kc98} and \citet{ng01}
investigated color-magnitude and metallicity-magnitude relations of
cluster elliptical galaxies.  This is extended to dwarf spheroidals,
$M_{B}\sim -10$, by \citet{ny04}.  \citet{k96}, \citet{spf01} and
\citet{ongy04} consider chemical evolution in spiral galaxies, a part of
which should be identified as damped Ly-$\alpha$ systems.  Some of them
found good agreement with observations.

Pioneering work taking into account chemical enrichment due to SNe Ia in
a SA model was carried out by \citet{t99} and \citet{tk99}.  In those
papers, they picked out averaged and individual formation histories of
dark halos, and then, assuming the closed-box chemical evolution, they
followed star formation and chemical enrichment histories.  Thus, while
they took into account merging histories of galaxies, the outflow of gas
caused by SNe or the supernova (SN) feedback that has been realized to
be an primarily important process in galaxy formation, was not
considered.  As shown by \citet{kc98} and \citet{ng01}, SN feedback
significantly affects chemical enrichment, at least due to SNe II.  This
suggests that the closed-box model has limitations in analyzing chemical
evolution in a realistic situation.

As a complementary approach to SA modeling, chemo-dynamical simulations
including SNe Ia have been developed \citep[e.g.,][]{rvn96, b99, k01,
lpc02, nm03, k03, oefj05}.  Although this approach has an advantage in resolving
spatial structure, because of the limitation of numerical resolution and
numerical techniques themselves, further improvements are still required
\citep{ojeqf03}.  On the other hand, the SA model is free from numerical
effects and limitations.  This is a great advantage for understanding global properties
of galaxy formation because in principle the SA model can investigate
from dwarf galaxies to galaxy clusters simultaneously.

In this paper, we construct a fully self-consistent treatment of
chemical enrichment with a SA model.  The basic model is the Mitaka
model presented by \citet{ny04}, which includes a Monte Carlo
realization of the merging histories of dark halos, radiative gas
cooling, quiescent star formation and starbursts, mergers of galaxies,
chemical enrichment assuming the IR approximation, size estimation of
galaxies taking into account the dynamical response to gas removal
during starbursts and stellar population synthesis.  As a first attempt,
we assume that all SNe Ia have the same lifetime.  This is, of course, a
rather simplified model.  In reality the lifetimes of SNe Ia are
considered to have a broad distribution, and there is even a claim that
low-metallicity environments inhibit SNe Ia \citep{ktnhk98}.  Our model,
however, has an essential characteristic of SNe Ia, that is, the time
lag between star formation and the explosion.  It enables us to see how
such a delayed explosion of SNe Ia affects the chemical enrichment and
abundance pattern in galaxies.  Using this model, we focus on the
chemical enrichment in Milky Way (MW)-like galaxies.  This has a particularly
meaning because there are many non-trivial effects on the chemical
enrichment in the hierarchical formation of galaxies.  Massive galaxies
such as the MW form not only via gas cooling but via mergers of
pre-galactic sub-clumps with stronger efficiencies of the SN feedback
than those in massive galaxies residing in deeper gravitational
potential wells.  { Thus, the main purpose of this paper is to
compare the SA model with a monolithic collapse model of \citet{ytn96}.
This will enable us to see how the hierarchical formation process
affects galactic metal enrichment due to SNe Ia based on the CDM model.}
Further comparison to other objects will be done in subsequent papers.

This paper is outlined as follows.  In \S2 we describe our SA model.  In
\S3 we provide a detailed prescription of the chemical enrichment due to
SNe Ia.  In \S4 we show the luminosity function of galaxies that should
be compared with that in the Local Group, and properties of MW-like
galaxies.  In \S5 we compare our model with observations in the
[O/Fe]--[Fe/H] plane and the [Fe/H] distribution in a statistical
sense. In \S6 we briefly discuss individual galaxies.  In \S7 we
investigate parameter dependences of the main results.  In \S8 we
provide a summary and conclusions.

\section{SEMI-ANALYTIC MODEL}
Here we briefly describe how to form galaxies in our model, which is
based on the Mitaka model \citep{ny04}.  In a CDM universe, dark matter
halos cluster gravitationally and grow in mass through their mergers,
depending on the adopted power spectrum of the initial density
fluctuations.  In each of the merged dark halos, radiative gas cooling,
star formation, and gas reheating by supernovae occur.  The cooled dense
gas and stars constitute {\it galaxies}.  These galaxies sometimes merge
together in a common dark halo, and then more massive galaxies form.
During these processes, chemical enrichment due to SNe II is solved by
assuming the IR approximation, and the enrichment due to SNe Ia by using
the past star formation histories of individual galaxies (see the next
section).  Repeating these processes, galaxies form and evolve to the
present epoch.  The Mitaka model reproduces many observations such as
luminosity functions, cold gas fraction, sizes of disks and spheroidals,
surface brightnesses, the Faber-Jackson relation, mass-to-light ratio
and faint galaxy number counts.  The details, apart from chemical
enrichment due to SNe Ia are found in \citet{ny04}.

Throughout this paper, we consider only a $\Lambda$-dominated CDM
($\Lambda$CDM) cosmology with $\Omega_{0}=0.3, \Omega_{\Lambda}=0.7,
h=0.7, \sigma_{8}=0.9$ and $\Omega_{\rm b}=0.02h^{-2}$, and a power
spectrum given by \citet{s95}, which takes into account the effects of
baryon.

The merging histories of dark halos are calculated by using a method
given by \citet{sk99}.  The timestep is $\Delta z=0.06(1+z)$, which
corresponds to the dynamical timescale of dark halos collapsing at $z$.
Only dark halos with circular velocity larger than $V_{\rm low}=40$
km~s$^{-1}$ are regarded as isolated halos.  While the value of $V_{\rm
low}$ hardly affects the chemical evolution in MW-like galaxies that we
consider in this paper, we will compare the luminosity function of
galaxies in the Local Group (LG) with the model luminosity functions
with various $V_{\rm low}$ in Section \ref{sec:setting}.  The circular
velocities of {\it root} halos, which are halos at $z=0$, is set to be
$V_{\rm circ}=220$ km~s$^{-1}$, which is nearly equal to the rotation
speed of the MW.

The cycle among baryonic components in dark halos is computed as
follows.  Diffuse baryonic gas is shock-heated to the virial temperature
estimated from the depth of the gravitational potential well of its host
dark halo when it collapses.  This is called hot gas, whose mass is
$M_{\rm hot}$.  The hot gas dissipates its energy and cools by radiation
only for halos with $V_{\rm circ}\leq V_{\rm cut}=250$ km~s$^{-1}$, to
avoid formation of unphysically giant galaxies, in each timestep.  The
cooling rate is estimated by using the metallicity-dependent cooling
functions given by \citet{sd93}.  This cooled gas concentrates in the
vicinity of the center of the dark halo and constitutes a galactic disk.
The mass of cold gas is $M_{\rm cold}$.  Disk stars form by consuming
the cold gas with a star formation timescale $\tau_{*}$.  In proportion
to the star formation rate $\psi=M_{\rm cold}/\tau_{*}$, massive stars
explode as SNe II.  The energy released by SNe II reheats the cold gas
and a part of the cold gas is absorbed into the hot gas (SN feedback).
The reheating rate is given by $\dot{M}_{\rm reheat}=\beta\psi$, where
$\beta=(V_{\rm disk}/V_{\rm hot})^{-\alpha_{\rm hot}}$ and $V_{\rm hot}$
and $\alpha_{\rm hot}$ are free parameters.  In this paper the star
formation timescale is assumed to be $\tau_{*}^{0}(1+\beta)$, where
$\tau_{*}^{0}$ is a free parameter, which is called the constant star
formation model (CSF) in \citet{ny04}.  Because $\beta$ is much larger
than unity in low mass halos owing to their shallower gravitational
potential wells, most of cold gas is reheated and only a small fraction
of it turns into stars.  The above processes are computed during each
timestep.

When a dark halo collapses by merging two or more dark halos, the hot
gas components in the progenitor halos are immediately merged and
constitute a hot gas component in the new dark halo.  In contrast,
galaxies do not merge together immediately. At first we define a central
galaxy in the newly collapsing dark halo as the central galaxy in the
most massive progenitor halo, and the rest of the galaxies are regarded
as satellite galaxies.  Two modes of galaxy mergers are considered.  One
is a merger between a central and satellite galaxies, in which a
satellite falls into a central galaxy after losing its energy by
dynamical friction \citep{bt87}.  Another is that between satellite
galaxies by random collisions \citep{mh97}.  We estimate the timescales
of the dynamical friction and the random collisions for all satellites.
Satellites with sufficiently short merger timescales compared with the
timescale of subsequent collapse of the dark halo merge with a central
or other satellite galaxies.  If the mass ratio of stars and cold gas of
a smaller galaxy to a larger one, $f=m_{\rm small}/m_{\rm large}$, is
larger than $f_{\rm bulge}$, a starburst occurs with SN feedback and all
stars go into the bulge component.  Then no cold gas remains.  Otherwise
the smaller satellite is simply absorbed into the disk component of the
central or larger satellite without any activities of star formation.
We adopt $f_{\rm bulge}=0.4$.

In each timestep, the disk size is estimated as a rotationally supported
disk under the assumption of specific angular momentum conservation for
cooling hot gas.  Initially, hot gas is assumed to have the same
specific angular momentum as that of its host dark halo.  The angular
momentum of dark halos has a log-normal distribution in terms of the
so-called dimensionless spin parameter.  If the estimated disk size is
larger than before, we renew the size.  At the same time, we also set
the disk rotation velocity to be the same as $V_{\rm circ}$ of the dark
halo.  When a major merger occurs, the bulge size is estimated by
assuming energy conservation.  Then a starburst occurs.  Because the
mass of the galaxy after the starburst is different from that before, we
consider the dynamical response to the mass loss of the size and
velocity dispersion.  To do this, we use the formalism of the dynamical
response taking into account the underlying dark matter potential developed
by \citet{ny03}.  This process is required to reproduce observed
characteristics of dwarf spheroidals such as surface brightness and
velocity dispersion.

Through the above processes, we obtain star formation histories of
individual galaxies.  From those we estimate the luminosities of
galaxies using a stellar population synthesis technique.  We adopt the
simple stellar populations given by \citet{ka97}.

These procedures are the same as those in \citet{ny04}.  Readers who
are interested in the details will find them in that paper.  The key
parameters of the model are tabulated in Table \ref{tab:param}.

\begin{deluxetable}{cccccc}
\tabletypesize{\scriptsize}
\tablecaption{Model parameters\label{tab:param}}
\tablewidth{0pt}
\tablehead{
\colhead{$V_{\rm hot}$} & \colhead{$\alpha_{\rm hot}$} &
 \colhead{$\tau_{*}^{0}$} & \colhead{$y_{\rm II,O}$} &
 \colhead{$y_{\rm II,Fe}$} & \colhead{$y_{\rm Ia,Fe}$}}
\startdata
150 km~s$^{-1}$ & 4 & 1.3 Gyr & $7.19\times 10^{-3}$ & $3.40\times
 10^{-4}$ & $6.67\times 10^{-4}$\\
\enddata

\tablecomments{$V_{\rm hot}$, $\alpha_{\rm hot}$ and $\tau_{*}^{0}$ is
determined by matching the luminosity functions of galaxies and the mass
fraction of cold gas in spiral galaxies with observations.  See
\citet{ny04} for details.  The above chemical yields are the same as
those adopted by \citet{ytn96}.  In the case of Salpeter's IMF
\citep{s55} with a slope of 1.35, according to their paper, the lower
and upper mass limits correspond to 0.03$M_{\odot}$ and 50$M_{\odot}$,
respectively.  The so-called ``A'' parameter determining binary
fractions is 0.015-0.055.}

\end{deluxetable}

\section{CHEMICAL ENRICHMENT}\label{sec:chem}
The chemical enrichment is treated in a similar manner to \citet{ytn96},
in which they considered chemical enrichment of oxygen (O) and iron (Fe)
due to both SNe II and SNe Ia using the infall model in the framework of
monolithic collapse \citep{ayt92}.  In their paper, the basic equations
describing the gas fraction, $f_{g}$, and metallicities for the $i$-th
elements, $Z_{i}$, are written as
\begin{eqnarray}
\frac{df_{g}}{dt}&=&-\alpha \psi(t)+A(t),\\
\frac{d(Z_{i}f_{g})}{dt}&=&-\alpha
 Z_{i}(t)\psi(t)+
 Z_{A,i}A(t)
 +\alpha y_{{\rm II},i}\psi(t)\nonumber\\
&&+\alpha y_{{\rm Ia},i}\int_{0}^{t}\psi(t-t_{\rm Ia})g(t_{\rm
 Ia})dt_{\rm Ia},
\end{eqnarray}
where $A(t)$ denotes the gas infall rate, $g(t_{\rm Ia})$ is a
distribution function of the lifetime of SNe Ia, $t_{\rm Ia}$, $\alpha$
is the locked-up mass fraction in low-mass stars and dead stellar
remnants and $y_{{\rm II},i}$ and $y_{{\rm Ia},i}$ are the chemical
yields of the $i$-th element produced by SNe II and SNe Ia, respectively
(see equations 1 and 2 in their paper).  The star formation rate $\psi$
is assumed to obey the Schmidt-law \citep{s63}, 
\begin{equation}
 \psi(t)=\nu f_{g}^{k},
\end{equation}
where $\nu$ is a rate coefficient.  The infall term is assumed to be
\begin{equation}
 A(t)=\frac{(t/t_{\rm in})^{\gamma}\exp(-t/t_{\rm in})}{t_{\rm
 in}\Gamma(\gamma+1,T_{G}/t_{\rm in})},
\end{equation}
where $t_{\rm in}$ is the infall timescale, $\Gamma(a,b)$ is the
incomplete gamma function and $T_{G}$ is the age of the Galaxy.  In the
followings, we compare our results with their $k=1$ model having $k=1$,
$(\alpha\nu)^{-1}=3.01$ Gyr, $t_{\rm in}=5$ Gyr, $\gamma=1$, and a
constant lifetime of SNe Ia, $g(t_{\rm Ia})=\delta^{D}(t_{\rm
Ia}-1.5{\rm Gyr})$.  This value of the lifetime of SNe Ia is consistent
with a theoretically estimated value using a model of \citet{lpc02},
which is based on \citet{gr83}, assuming Kennicutt's initial mass
function (IMF) of stars \citep{k83}.  We have estimated that the mean
lifetime is about 2.4 Gyr and the median about 1.8 Gyr.

In our model, because of the SN feedback, these equations are modified.
During disk star formation or a starburst in a timestep that corresponds
to $\Delta z=0.06(1+z)$, chemical enrichment due to SNe II is solved by
assuming the IR approximation,
\begin{eqnarray}
\frac{dM_{\rm cold}}{dt}&=&-(\alpha+\beta)\psi(t),\label{eqn:Mcold}\\
\frac{dM_{\rm hot}}{dt}&=&\beta\psi(t),\label{eqn:Mhot}\\
\frac{d(M_{\rm cold}Z_{{\rm cold},i})}{dt}&=&-(\alpha+\beta)Z_{{\rm
cold},i}\psi(t)+\alpha y_{{\rm II},i}\psi,\label{eqn:Zcold}\\
\frac{d(M_{\rm hot}Z_{{\rm hot},i})}{dt}&=&\beta Z_{{\rm cold},i}\psi,\label{eqn:Zhot}
\end{eqnarray}
where $Z_{{\rm cold},i}$ and $Z_{{\rm hot},i}$ are metallicities for
$i$-th element of cold and hot gas, respectively.  The star formation
rate $\psi$ is given by $M_{\rm cold}/\tau_{*}$.  Equations
(\ref{eqn:Mcold})-(\ref{eqn:Zhot}) do not have infall terms because we
add the mass of hot gas that should cool during a timestep $\Delta z$
to the mass of cold gas at the beginning of each timestep.  The
timestep is sufficiently short to describe smooth gas accretion.
Therefore no infall term is required as an
approximation.  The mixture of merging histories of dark halos, gas
cooling and SN feedback should correspond to the infall process with a
long timescale, $\sim 5$Gyr.  The roles of the physical processes are
clarified in Section 6.  In addition to this process, metals released
from SNe Ia are put into the cold gas.  The mass of $i$-th element
released during a timestep [$t_{j-1},t_{j}$] is
\begin{equation}
 \Delta M_{i}=\alpha y_{{\rm Ia},i}\int_{t_{j-1}}^{t_{j}}\psi(t-t_{{\rm Ia}})dt.
\end{equation}
In order to estimate this quantity, star formation histories of
individual galaxies are stored.  Again, we would like to stress that
the above procedures of chemical enrichment due to both SNe Ia and
II are computed in each timestep.

Since one of the main purposes of this study is to compare our results
with a monolithic cloud model given by \citet{ytn96}, we adopt the same
chemical yields empirically determined by their analysis, which are also
tabulated in Table \ref{tab:param}.  Thus we also impose two
nucleosynthesis constraints they adopted: $y_{\rm II,O}/y_{\rm II,Fe}=$
const., and $y_{\rm Ia,O}=0$.  { Returned mass from SNe Ia is simply
treated together with that from massive stars, that is, it is included
in a parameter for the instantaneous recycling, $1-\alpha$.}  For
simplicity we do not take into account the following effects that might
affect chemical evolution of galaxies.  { One is energy feedback due
to SNe Ia, which was investigated in detail in \citet{kg03a, kg03b}.}
{ Second} is inhomogeneous star formation in a gas cloud considering
propagation of star formation sites, which is another way to extend the
simple traditional galaxy evolution model \citep{tsy99, iw99}.  Although
it will probably provide a broader distribution in metal abundances, we
concentrate, in this paper, on how the hierarchical formation process
affects the chemical enrichment.  { Third} is the metallicity effect
that a low metallicity environment inhibits the explosion of SNe Ia
\citep{ktnhk98}, which might affect the estimation of $t_{\rm Ia}$.
These effects would be worth considering in future.

\section{luminosity function and global characteristics of MW-like galaxies}\label{sec:setting} 
The model we use here is the same as that in \citet{ny04} except for
the chemical enrichment process, which do not affect any results in that
paper.  Therefore we do not repeat showing the same figures as shown in
that paper.

In the following, we pick out galaxies in dark halos of $V_{\rm
circ}=220$ km~s$^{-1}$ whose central galaxies are identified as spiral
galaxies with $B$-band bulge-to-total luminosity ratio smaller than 0.4
\citep{bcf96} and with $-22.1\leq M_{I}-5\log h\leq-21.6$ \citep{sp99}
and disk rotation velocities $210\leq V_{\rm disk}/$km s$^{-1}\leq 230$
\citep{blbcf02a}.  Hereafter, we call a dark halo satisfying the above
criteria a LG-halo.  We realize 100 LG-halos and average over the
realization.

\begin{figure}
\plotone{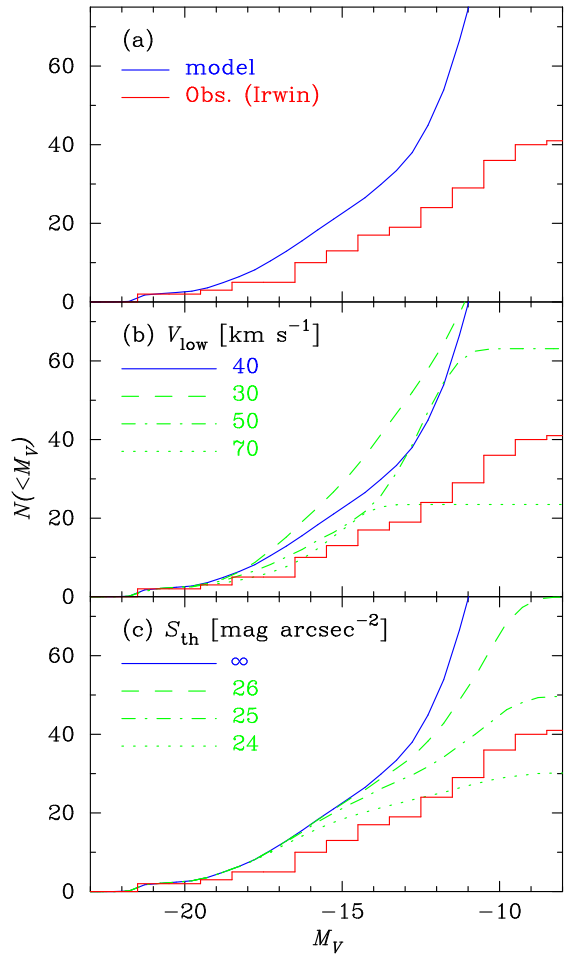}

\caption{{ Cumulative} Luminosity function in the LG.  (a) The solid
line indicates the reference model.  The histogram denotes observed
luminosity function in the LG compiled by M. Irwin.  (b) Dependence on
$V_{\rm low}$.  The solid line is the same as model in (a), the
reference model of $V_{\rm low}=40$ km~s$^{-1}$.  The dashed, dot-dashed
and dotted lines indicate variants of the reference model with $V_{\rm
low}=30, 50$ and 70 km~s$^{-1}$.  (c) Dependence on $S_{\rm th}$.  The
solid line is the same as model in (a), the reference model of $S_{\rm
th}=\infty$.  The dashed, dot-dashed and dotted lines indicate variants
of the reference model with $S_{\rm th}=26, 25$ and 24 mag
arcsec$^{-2}$, respectively.  }

\label{fig:lf}
\end{figure}

Before entering the details, we see the characteristics of MW-like
galaxies and their satellites.  In Figure \ref{fig:lf}a we show a
luminosity function of galaxies in LG-halos.  The solid line indicates
the model luminosity function and the histogram the observed one
compiled by M. Irwin\footnote{See
http://www.ast.cam.ac.uk/\~{}mike/local\_members.html}.  Because the
observed data are for the MW and M31 groups, luminosity functions for
models are simply doubled to be compared with the observed data.

Our model shows a broad agreement with the observed
luminosity function in the LG at $M_{B}\la -15$, although at fainter
magnitude too many galaxies are predicted.  While the observed
luminosity functions actually depend on the definition of the Local
Group, our conclusion about the chemical evolution of the MW-like
galaxies does not change as shown below.  The detailed discussion is
given by \citet{bflbc02b}.

Such overabundance of low luminosity galaxies is easily eliminated by
taking into account some physical or observational effects as shown by
many authors.  One of the important effects is photoionization due to
ultraviolet (UV) background or reionization \citep[e.g.,][]{ngs99, ng01,
s02, blbcf02a, bflbc02b, bfbcl03}.  Here we show the effect of the Jeans
mass increasing due to reionization in a simple way by changing $V_{\rm
low}$.  Note that in reality this effect should turn on only after the
epoch of reionization.  Indeed, this simple model is similar to
\citet{kwg93}, in which they assumed that gas cooling is prohibited in
dark halos with $V_{\rm circ}\leq 150$ km~s$^{-1}$ at redshifts between
1.5 and 5 to be consistent with the LG luminosity function in the
standard CDM model.  In Figure \ref{fig:lf}b, we show luminosity
functions with different $V_{\rm low}$.  As shown in the panel the
solid, dashed, dot-dashed and dotted lines indicate $V_{\rm low}=$ 40
(the reference model), 30, 50 and 70 km~s$^{-1}$, respectively.
Evidently large $V_{\rm low}$ inhibits the formation of dwarf galaxies.
Another effect is incompleteness in the observations.  \citet{ny04}
predict that there are many dwarf galaxies with surface brightnesses too
low to detect.  In the bottom panel, Figure \ref{fig:lf}c, we show
luminosity functions with different threshold surface brightnesses,
$S_{\rm th}$, which determines the lowest surface brightness of galaxies
to be detected.  The solid, dashed, dot-dashed and dotted lines indicate
$S_{\rm th}=\infty$, 26, 25 and 24 mag~arcsec$^{-2}$, respectively,
where surface brightness is defined as that within the effective radius.
{ Although the lowest one is about 26 mag arcsec$^{-2}$ \citep{m98},
clearly observations are not a surface brightness-complete sample.} {
As shown in \citet{ny04}, the lowest surface brightness of model
galaxies reaches about 30 mag arcsec$^{-2}$ in $B$-band at $M_{B}\simeq
-10$.} In addition, some of these low surface brightness galaxies might
be disrupted by tidal interaction with other galaxies
\citep[e.g.,][]{mibild03}.  These figures suggest that the so-called
satellite-galaxy problem might not be so serious, at least for luminous
objects.  Anyway, changing $V_{\rm low}$ and $S_{\rm th}$ hardly affects
our results on chemical enrichment of the MW-like galaxies.

\begin{figure}
\plotone{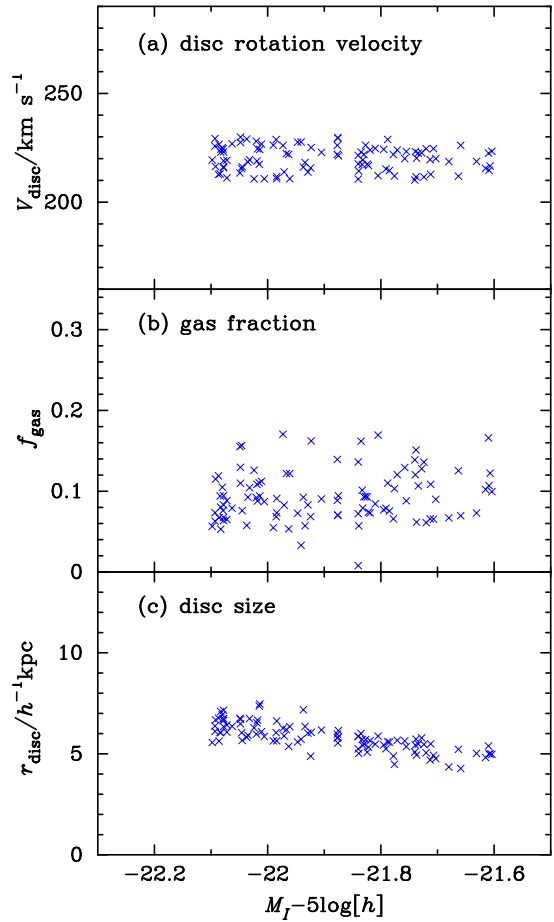}

\caption{Characteristics of MW-like galaxies.  The crosses indicate each
 MW-like galaxy.  (a) Disk rotation velocity $V_{\rm disk}$.  (b) Cold
 gas fraction.  (c) Effective radius of disks.}

\label{fig:other}
\end{figure}

In Figure \ref{fig:other}, we show the disk rotation velocity, $V_{\rm
disk}$, cold gas fraction, $f_{\rm gas}$, and effective radius of the
disk, $r_{\rm disk}$, of the MW-like galaxies.  The crosses denote
individual MW-like galaxies.  Since the disk rotation velocity is
adjusted to be the circular velocity of the host dark halo only when the
mass of disk increases, it is not always the same as $V_{\rm circ}$,
which is chosen to be 220 km~s$^{-1}$ at $z=0$.  The rotation velocity
obtained is distributed around the value 220 km~s$^{-1}$ that is similar
to the observed rotation velocity of the Galactic disk.  The cold gas
fraction is about 10-15\% in mass, and the effective radius is around
7$h^{-1}$kpc.  These are very similar to the observed properties of the
MW.  Thus our MW-like galaxies are good probes enough to investigate the
chemical enrichment in the MW.

\section{COMPARISON WITH SOLAR NEIGHBORHOOD AND BULGE STARS}
In this model we treat bulge and disk stars separately, so hereafter we
show results for both separately, keeping in mind there are still large
uncertainties in observation of bulge stars.  In the followings, we
compare our results not only with a theoretical infall model given by
\citet{ytn96} but also with stars in our Galaxy.  { We focus on the
difference between the hierarchical and monolithic models, rather than
comparisons with observations.}  We assume that our Galaxy is a
representative, typical sample of galaxies of MW-like galaxies under
consideration.  Future observations will make clear whether this
assumption is valid or not.

Firstly, in this section, we show mean properties of MW-like galaxies.
Properties of individual galaxies are discussed in \S\ref{sec:ind}.

\begin{figure*}
\plotone{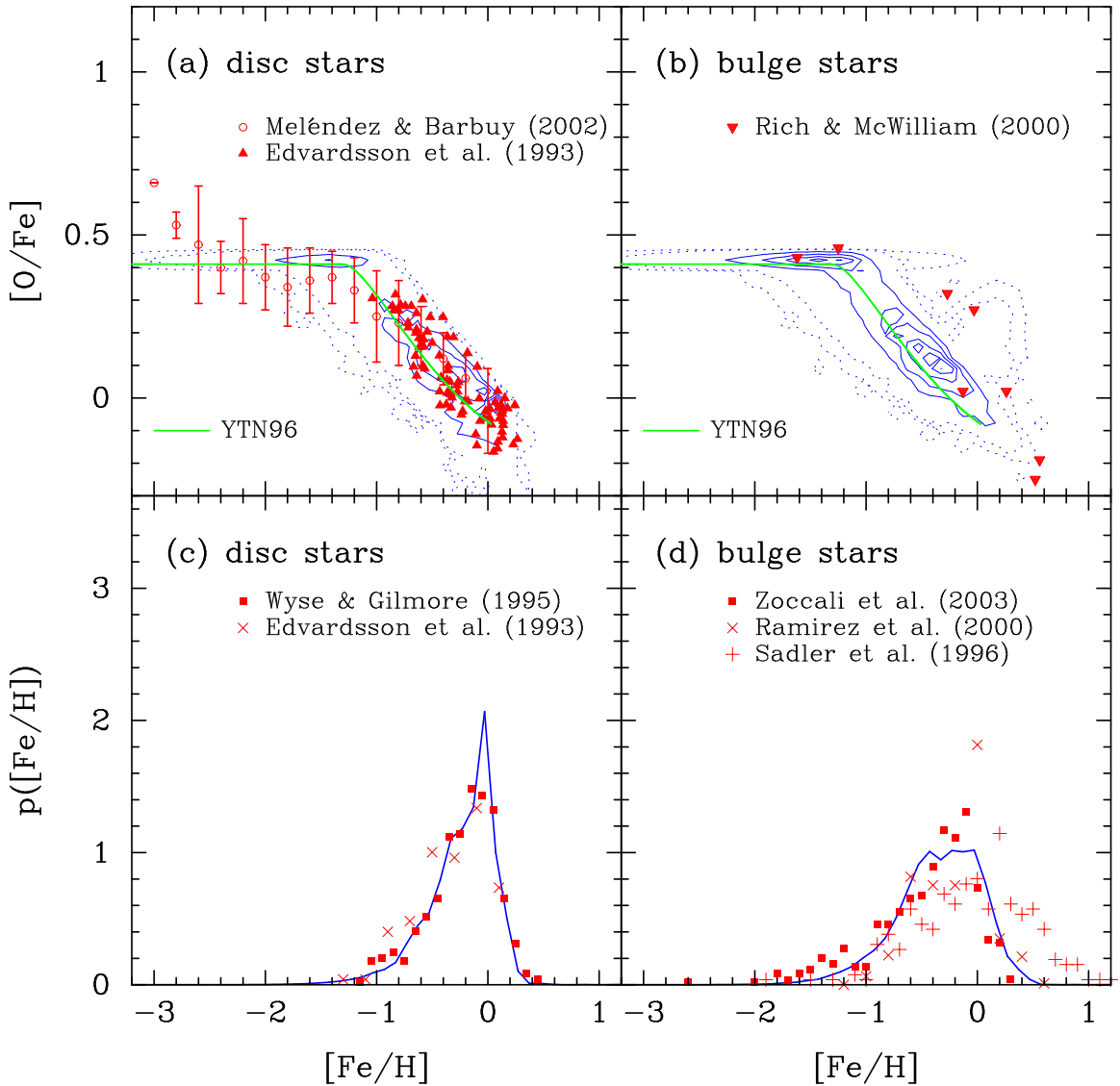}

\caption{(a) [O/Fe] distribution against [Fe/H] for disk stars.  The
levels of contours drawn by the thin solid and dashed lines indicate
0.5, 0.4, 0.3, 0.2 and 0.1, and 0.02 and 0.005 times the largest number
of stars in grids, respectively.  The solid curve indicates a chemical
enrichment model based on a monolithic cloud collapse model given by
\citet{ytn96}.  The circles with error bars and filled triangles denote
observations by \citet{mb02} and \citet{eaglnt93}, respectively.  (b)
[O/Fe] distribution against [Fe/H] for bulge stars.  The levels of
contours and the solid curve are the same as the panel (a). The filled
inverse triangles denote observational data by \citet{rm00}.  (c) {
Iron MDF} for disk stars.  The solid line indicates the model prediction
for disk stars.  The filled squares and crosses denote observations by
\citet{wg95} and \citet{eaglnt93}, respectively.  (d) { Iron MDF} for
bulge stars.  The solid line indicates the model prediction for bulge
stars.  The filled squares, crosses and pluses denote observations by
\citet{z03}, \citet{rsfd00} and \citet{srt96}, respectively.  }

\label{fig:fid}
\end{figure*}

In Figure \ref{fig:fid}a we show the distribution of disk stars in the
[O/Fe]--[Fe/H] plane by contours.  By tracking star formation histories
and metallicity evolutions, we obtain distributions of stars in the
plane for individual galaxies.  The figure shows the average of 100
MW-like galaxies.  The levels of contours drawn by the thin solid and
dashed lines indicate 0.5, 0.4, 0.3, 0.2 and 0.1, and 0.02 and 0.005
times the largest number of stars in grids, respectively.  The solid
lines indicate the $k=1$ model given by \citet{ytn96} (see Tables 1 and
2 in their paper for details).  The crosses and filled triangles are
observed data for solar-neighborhood stars given by \citet{mb02} and
\citet{eaglnt93}, respectively.  As is clearly shown in this panel, our
model reproduce well the observations of disk stars including their
dispersion.  The apparent discrepancy at [Fe/H]$\la-2.5$ is caused by
the first nucleosynthesis constraint, $y_{\rm II,O}/y_{\rm II,Fe}=$
const. for all SNe II (see Section \ref{sec:chem}).  At the break point
the model prediction gives slightly higher [O/Fe] than observations.
Probably if the assumption of a constant $t_{\rm Ia}$ were relaxed, the
abundance ratio at there would decrease as suggested by \citet{ytn96}.
{ The dispersion for the model comes from dispersion in a single
galaxy and that for different samples.  In the next section we show the
former effect by seeing some individual galaxies.}

It should be noted that a small radial abundance gradients of metals in
the galactic disk are suggested by observations, $\sim -0.085$ dex
kpc$^{-1}$ for iron and $\sim -0.07$ dex kpc$^{-1}$ for oxygen
\citep{m01}.  This might cause a systematic difference in the
metallicity distribution when averaged over the whole disk.  In that
case, we will need to change the values of chemical yields, which
are empirically determined by \citet{ytn96} to produce an agreement with
the observations for solar-neighborhood stars.

Bulge stars are also distributed like disk stars in the [O/Fe]--[Fe/H]
plane but the range is slightly different as shown in Figure
\ref{fig:fid}b.  In particular, although the fraction is low, there are
many oxygen-enhanced stars at [Fe/H]$\simeq 0$.  Interestingly,
\citet{rm00} found that [O/Fe] for bulge stars, shown by filled inverse
triangles, has a broadly similar trend with [Fe/H] to disk stars, but
some of them show oxygen-enhancement, [O/Fe]$\simeq 0.3$ even at
[Fe/H]$\simeq 0$.  Recent high-resolution spectroscopic observations of
bulge stars by \citet{mrs03} reveal a similar trend of [Mn/Fe] to that
for disk stars.  While they observed less than 10 stars, their
observations may support our model.  Clearly more samples will provide
important constraints on bulge formation.  Using monolithic collapse
models, \citet{mb90} predicted more oxygen-enhanced bulge stars.  To
obtain this, they assumed higher star formation rates and shorter infall
timescales for bulge stars than those for disk stars.  In our model,
some bulge stars are also formed with a very short timescale, that is,
starbursts.  The difference we predict between disk and bulge stars,
however, is smaller presumably because not all bulge stars are formed by
starbursts and some have been formed in progenitor galaxies.  Another
possibility to form oxygen-enhanced stars is a variation of the IMF.
Recently the IMF for starbursts is suggested to be top-heavy compared to
the IMF for the disk star formation \citep{baugh05, nlbfc05, nlobfc05},
{ which means that much more SNe II must explode than expected for
Salpeter-like IMFs.}  Because $\alpha$-elements { from SNe II} are
likely to be produced more, metal-rich bulge stars will tend to be
placed at [O/Fe]$\ga 0$.

In Figures \ref{fig:fid}c and d, we show the { iron MDF} for disk and
bulge stars, respectively.  The solid lines denote the model prediction.
The filled squares and crosses in the panel (c) and the filled squares,
crosses and pluses in the panel (d) indicate observed [Fe/H]
distributions given by \citet{wg95} and \citet{eaglnt93} for disk stars
and \citet{z03}, \citet{rsfd00} and \citet{srt96} for bulge stars,
respectively.  All of these distributions are normalized for the
integrate over the whole range to be unity.  The distribution for disk
stars shows an excellent agreement with the observed distributions.
Agreements in abundance ratio and { MDFs} are quite encouraging to
model the formation of the galactic disk in the framework of SA models.
{ For bulge stars, there remains discrepancy among observations
themselves.  Our model is consistent only with the data from \citet{z03}
and \citet{rsfd00}.  The data from \citet{srt96} shows more metal-rich
stars than the other two and our model. }

While the predicted MDF for bulge stars is rather similar to that for
disk stars, the width is clearly wider than that of disk stars.  The
observations of bulge stars show a similar tendency to the model,
although there seem to be still uncertainties in observations outside
the range $-1\la$[Fe/H]$\la 0$.  Keeping in mind differences between
observational data, the model prediction for bulge stars also agrees
well with the observed data.  In another approach, a recent
chemo-dynamical simulation of the galactic bulge by \citet{nm03} also
predicts distribution of stars in [O/Fe]-[Fe/H] plane and in number
against [Fe/H].  In their results, there are more oxygen-enhanced stars
than in the observation by \citet{rm00} and our prediction.  At the same
time, their derived [Fe/H] distribution is similar to ours, although the
number of high [Fe/H] stars is a little larger.  Thus, although more
data will be required to constrain the model of bulge formation, we can
say that both our SA model and the recent numerical simulation broadly
agree with the current observations.

\begin{figure}

\plotone{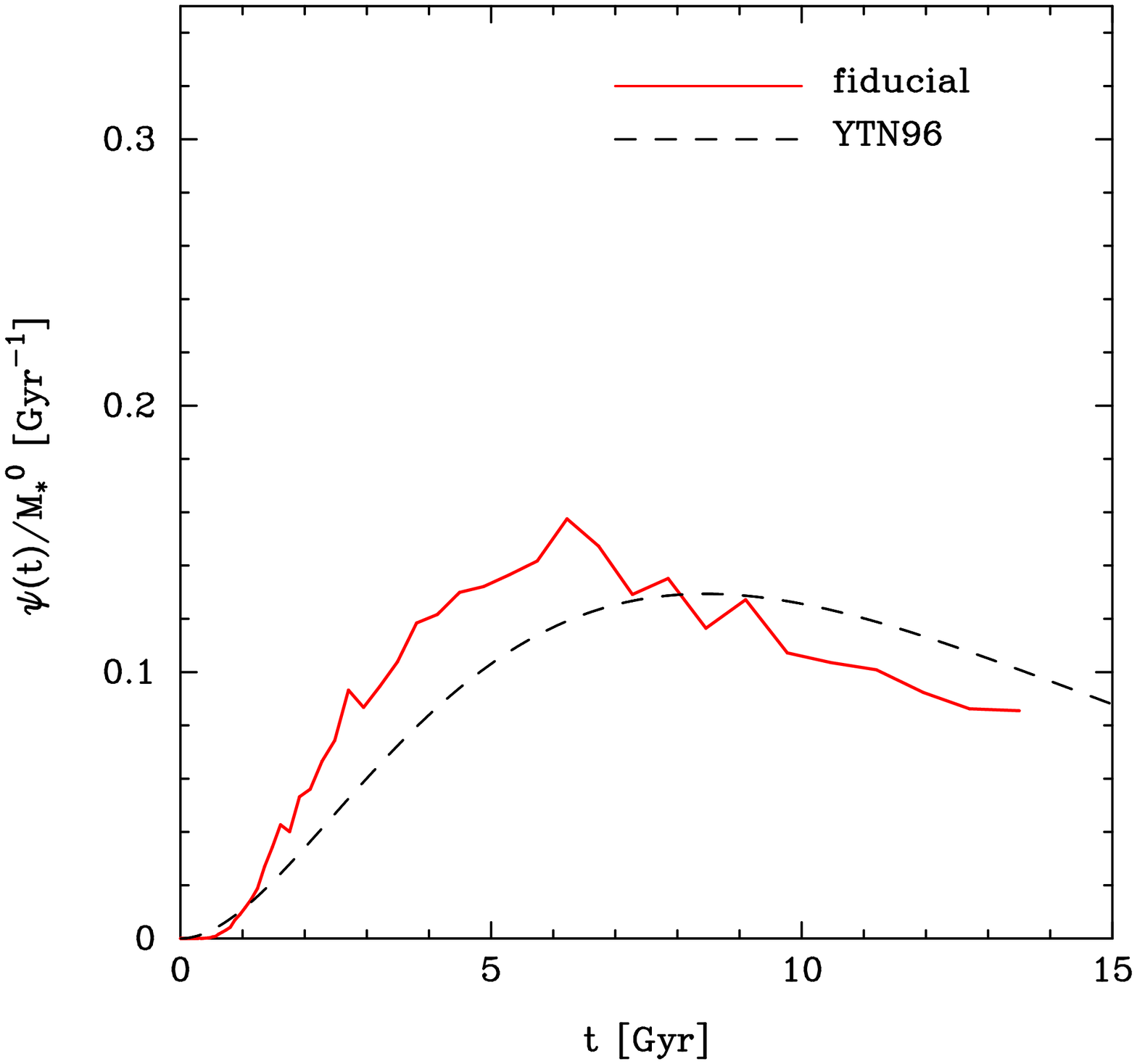}

\caption{Star formation histories.  The solid line indicates the
averaged star formation history of MW-like galaxies.  The dashed line
denotes a history given by the infall model of \citet{ytn96}.  }

\label{fig:sfr1}
\end{figure}

Figure \ref{fig:sfr1} shows averaged star formation histories of MW-like
galaxies indicated by the solid line.  The infall model given by
\citet{ytn96} is indicated by the dashed line.  { Note that the SFR
for the infall model is plotted beyond the cosmic age given by the
cosmological parameters we adopt, $\simeq 13$ Gyr, because their model
is normalized at 15Gyr.}  The star formation rates are normalized by
stellar masses at 13Gyr.  { SFRs plotted are those averaged over each
timestep ($\Delta M_{*}/\alpha\Delta t$) and over all realization.}
Clearly galaxies in our SA model have similar star formation histories
to the monolithic collapse model's.  Hence the slow star formation in
the infall model, which is brought by long infall and star formation
timescales chosen by obtaining an agreement with observations, is
naturally explained by the mixture of physical processes including
merging histories of dark halos, gas cooling and SN feedback.  While a
history of each MW-like galaxy has a variety around the averaged
history, { as we will show in the next section,} we have found that
predicted metallicity distributions of disk stars are quite similar to
the averaged one shown in Figure \ref{fig:fid}.  Since bulge components
are formed by major mergers that are rare events, it should be noted
that the results on the bulge stars as just an average.  Future
observations of bulge stars in other galaxies will provide stronger
constraint on galaxy formation models based on the CDM cosmology.

Our results suggest that the formation of spiral galaxies is well
approximated by the traditional infall model because our model provides
similar results to those given by \citet{ytn96}.  This is consistent
with the result by \citet{bcf96}, in which they traced formation
histories of individual galaxies and found similar star formation
histories of spiral galaxies to the traditional infall model (see upper
panels of Figure 3 in their paper).  Note that recent infall models have
become more complicated, including multi-accretion of gas, in order to
reproduce observational results.  Moreover, to reproduce radial
gradients of metallicities, the models require SFRs dependent on the
radius of disks \citep[e.g.,][]{cmg97, fg03}.  Physically, the radial
gradients should be solved by computing complex dynamics of disk
formation.  This requires numerical simulations of galaxy formation
following dynamical evolution including the mixing process of gas and
metals.

\begin{figure}
\plotone{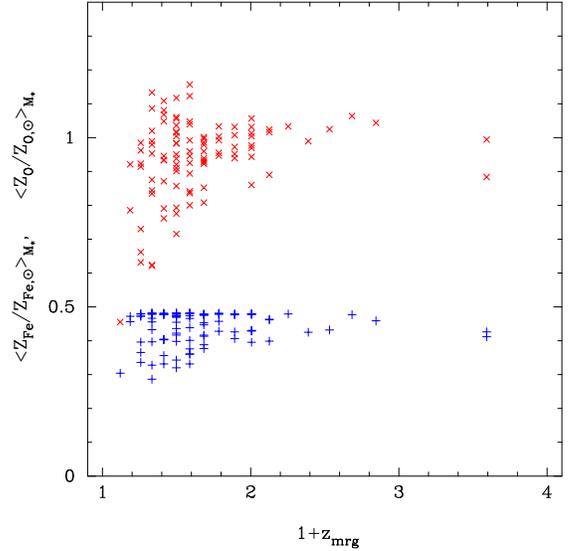}

\caption{Mass-weighted metallicities of O ({\it crosses}) and Fe ({\it
pluses}) for disk stars of individual MW-like galaxies against their
last major merger epochs, $z_{\rm mrg}$.}

\label{fig:Z_z}
\end{figure}

Finally we shortly discuss relations between averaged metallicities of
individual galaxies and their merging histories.  Figure \ref{fig:Z_z}
shows mass-weighted mean metallicities of disk stars for O ({\it
crosses}) and Fe ({\it pluses}) against their final major merger epochs,
$z_{\rm mrg}$.  Correlations of metallicities with $z_{\rm mrg}$ are
very weak for O and negligible for Fe.  Particularly, the Fe-metallicity
is almost independent of $z_{\rm mrg}$.  The O-metallicity only for
galaxies which experienced a major merger very recently is slightly
lower or has a larger scatter than others.  To make the reason clear,
however, we need more samples of high-z MW-like galaxies.  The current
number of high-z samples is still few for quantitative statistical
discussion.  At this stage, what we can say is that there is no clear
correlation between the metallicity and the final major merger epoch.

\begin{figure}
\plotone{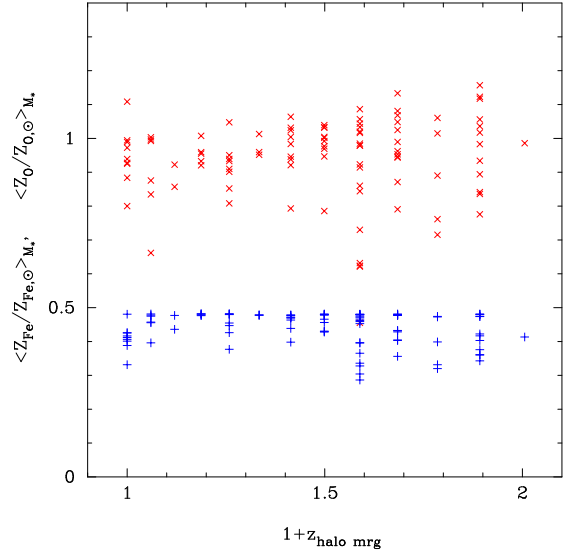}

\caption{Mass-weighted metallicities of O ({\it crosses}) and Fe ({\it
pluses}) for disk stars of individual galaxies against last major
merger epochs of their host halos, $z_{\rm halo~mrg}$.}

\label{fig:Z_zh}
\end{figure}

Figure \ref{fig:Z_zh} shows the same as the previous figure but for
final major merger epochs for their host halos, $z_{\rm halo~mrg}$.
Similarly, we do not see correlations between the O-metallicity and
$z_{\rm halo~mrg}$.  The negligible or weaker correlation with $z_{\rm
halo~mrg}$ than with $z_{\rm mrg}$ can be reasonable because the galaxy
merger does not directly reflect the halo merger and is lead by other
physical mechanisms, namely, dynamical friction and random collisions.

Thus, we conclude that the metallicity depends only weakly on the
formation history of galaxies and their host halos.  Instead, we can see
much stronger dependence of the metallicity on parameters of, for
example, SN feedback later.

\section{Individual galaxies}\label{sec:ind}
In the previous section, we showed mean properties of MW-like galaxies.
We here show results for two samples in order to see properties of
individual galaxies.

\begin{figure}
\plotone{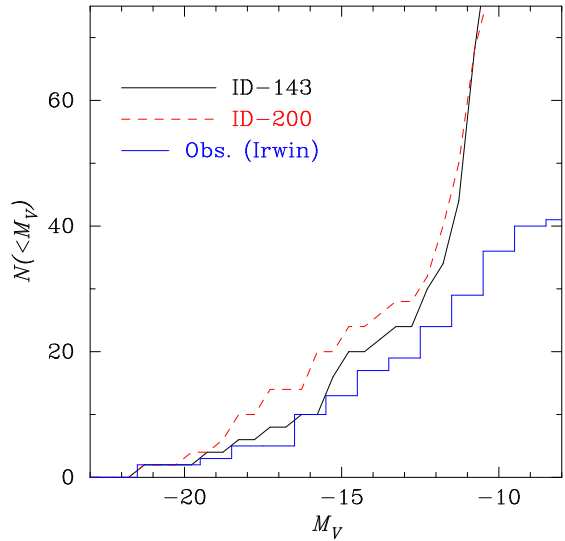}

\caption{Cumulative luminosity functions in LGs of ID-143 ({\it solie
line}) and ID-200 ({\it dashed line}).  No selection effect is taken
into account.  The histogram denotes the same observed luminosity
function shown in Figure \ref{fig:lf}.}

\label{fig:lf_ind}
\end{figure}

Figure \ref{fig:lf_ind} shows luminosity functions of galaxies in two
LG-halos hosting MW-like galaxies as central galaxies of ID-143 with
$z_{\rm mrg}=1.13$, $z_{\rm halo~mrg}=0.79$, $\langle Z_{\rm O}/Z_{\rm
O,\odot}\rangle=0.89$, and $\langle Z_{\rm Fe}/Z_{\rm
Fe,\odot}\rangle=0.40$, and ID-200 with $z_{\rm mrg}=0.59$, $z_{\rm
halo~mrg }=0.50$, $\langle Z_{\rm O}/Z_{\rm O,\odot}\rangle=0.99$, and
$\langle Z_{\rm Fe}/Z_{\rm Fe,\odot}\rangle=0.47$, respectively.  The
luminosity function in ID-143 halo agrees well with the observed one at
$M_{V}\la -15$, and that in ID-200 halo broadly agrees but slightly
overproduces galaxies at $M_{V}\sim -17$.  This slight discrepancy
between the models and the observations should substantially decrease
when taking into account some effects such as the selection effects
shown in \S\ref{sec:setting}.  We also find that difference between
samples is not significant.

\begin{figure}
\plotone{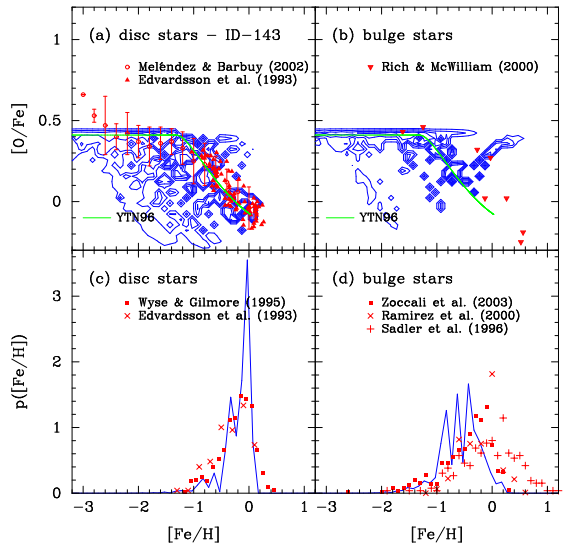}

\caption{The same as Figure \ref{fig:fid} but for a MW-like galaxy of
ID-143.  Note that, in order to enhance contribution of stars with small
fractions, the contour levels are equally divided in logarithmically.}

\label{fig:ri143}
\end{figure}

\begin{figure}
\plotone{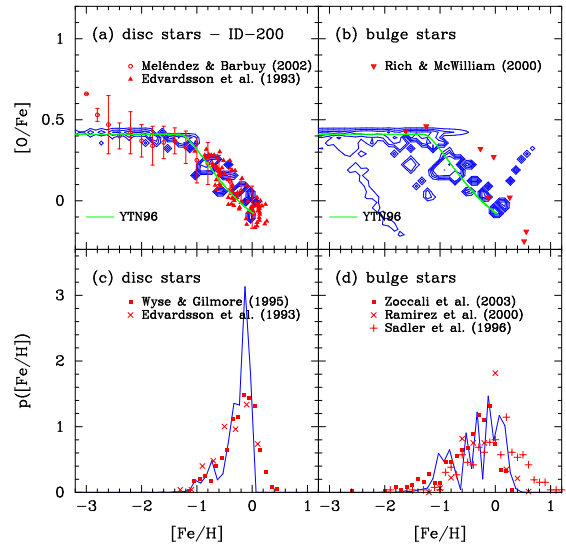}

\caption{The same as Figure \ref{fig:ri143} but for a MW-like galaxy of
ID-200.}

\label{fig:ri200}
\end{figure}

Figures \ref{fig:ri143} and \ref{fig:ri200} show the same figures as
Figure \ref{fig:fid}, that is, the distribution of disk and bulge stars
in [O/Fe]--[Fe/H] plane and MDFs for [Fe/H], but for individual galaxies
of ID-143 and ID-200, respectively.  In order to enhance the
contribution of stars with small fractions, the contour levels are not
linear but logarithmic.  As easily can be seen, some stars are located
in a low [O/Fe] region at low [Fe/H], particularly for ID-143.  Such
stars can be brought by smaller sub-clumps via minor mergers.  Because
low-mass galaxies with shallow gravitational potential wells suffer
strong SN feedback, the metal enrichment does not proceed significantly.
Therefore SNe Ia start to explode and lower [O/Fe] when [Fe/H] is still
low.  This effect shall be shown in Figure \ref{fig:snfb1} for a
stronger SN feedback case.  Although the number of such stars is small,
it might be discovered as relics of past minor merger histories.  Note
that, of course, this depends on the details of the process of minor
mergers.

In contrast to ID-143, the galaxy ID-200 does not have such low-[O/Fe],
low-[Fe/H] disk stars.  This is because minor mergers did not happen
after the last major merger and the disk grew in a similar way to the
infall model of monolithic collapse.  Merging histories of these
galaxies should provide insights for understanding the difference in the
stellar distribution in [O/Fe]-[Fe/H] plane.  For example, the last
merger epoch for this galaxy is $z_{\rm mrg}=0.59$, and for ID-143
galaxy with many minor mergers $z_{\rm mrg}=1.13$, which is earlier than
that of ID-200's.  Thus it should have been hard for ID-200 galaxy to
accrete small sub-clumps as minor mergers until present.

Similar trails can be seen for bulge stars both of ID-143 and ID-200
galaxies, in contrast to disk stars.  Because many stars in bulges
formed in disks firstly and then they were transferred to bulges by major
mergers, the trails in bulges should reflect minor-merger histories onto
disks before formation of bulges, or, last major mergers.  Therefore
these galaxies could have enough time to accrete sub-clumps until last
major mergers.

The widths of MDFs for both MW-like galaxies are slightly narrower than
the averaged one shown in Figure \ref{fig:fid}.  In our model we assume
that metals are well mixed in a galaxy.  In reality, however, the mixing
of metals in a galaxy would not be completed and therefore the MDF would
be broadened as observed.  If we take this effect of inhomogeneous metal
enrichment into account, the MDF will become wider.  Note that recent
observations of more than 10,000 solar-neighborhood stars by \citet{n04}
have suggested a narrower MDF than past observations.  Although we
should be careful in sample selection, it may suggest that the metal
mixing is efficient for disk stars.  This should be addressed in
future.

\section{PARAMETER DEPENDENCE}\label{sec:depend}
Here we investigate how the stellar metallicity depends on the
fundamental model parameters such as the SN feedback and star formation
timescale, as well as the lifetime of SNe Ia, in order to clarify how
physical processes such as star formation affect chemical enrichment in
galaxies.  { Note that adopted parameters below break the agreement
with observations.  We vary the values of the parameters just to see how
each physical process affects metal enrichment.}

\subsection{Lifetime of SNe Ia}
\begin{figure}

\plotone{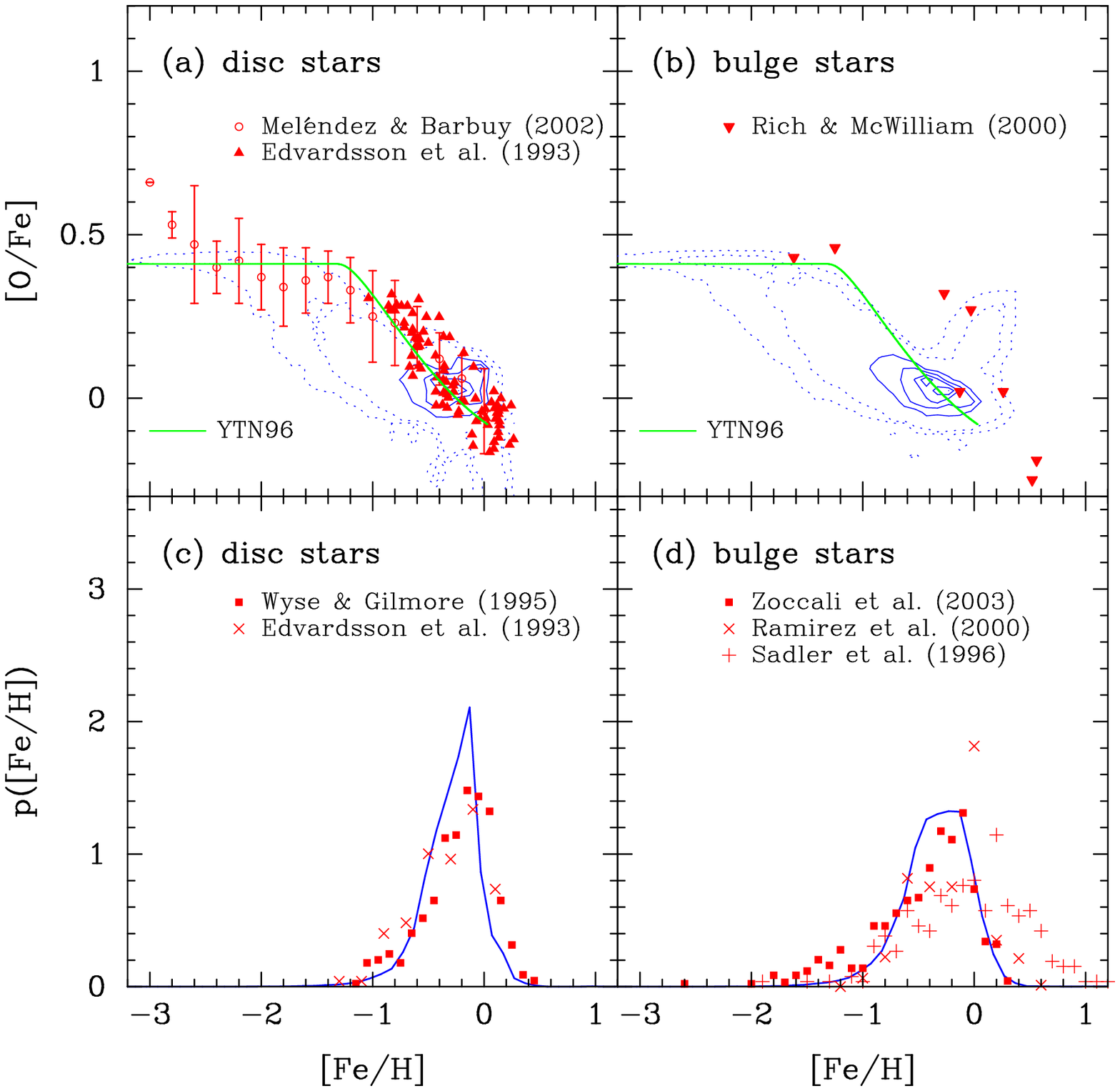}

\caption{The same as Figure \ref{fig:fid} but for $t_{\rm Ia}=0.5$ Gyr.}

\label{fig:tIa1}
\end{figure}

\begin{figure}

\plotone{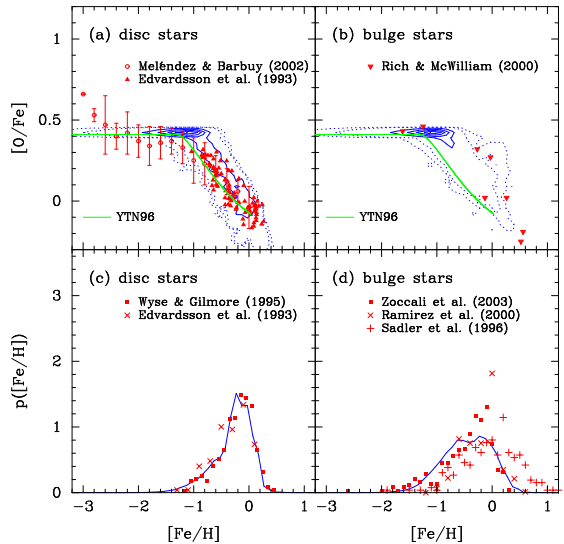}

\caption{The same as Figure \ref{fig:fid} but for $t_{\rm Ia}=3$ Gyr.}

\label{fig:tIa2}
\end{figure}

Firstly, we vary the lifetime of SNe Ia, $t_{\rm Ia}$.  In Figures
\ref{fig:tIa1} and \ref{fig:tIa2}, we show the same model as in Figure
\ref{fig:fid} but for $t_{\rm Ia}=0.5$ and 3 Gyr, respectively.  In the
case of the shorter $t_{\rm Ia}$, because enrichment due to SNe Ia
begins earlier, the break point in the [O/Fe]--[Fe/H] plane moves toward
lower [Fe/H].  On the other hand, in the longer $t_{\rm Ia}$ case, the
iron abundance initially increases only due to SNe II, so the break
point moves toward higher [Fe/H].  This dependence is basically the same
as that shown in analyses using the traditional model.  This also
suggests the similarity of formation of spiral galaxies between the
hierarchical and traditional models.

In the shorter $t_{\rm Ia}$ case, an interesting feature in abundance
ratio emerges there.  Bulge stars has a kind of U-shaped distribution
with a minimum at [Fe/H]$\sim-0.5$ in [O/Fe]--[Fe/H] plane.  In this
case, iron enrichment due to SNe Ia is almost completed at very high
redshift, at which time a large amount of cold gas remains in galaxies.
Hence most of the stars existing at present form in subsequent starburst
due to major mergers between such gas-rich systems.  Because starbursts
tends to make stars at its late phase with the abundance pattern of SNe
II, if the system is gas-rich, many oxygen-enhanced iron-rich stars form
during the starburst.  Thus this provides a constraint on the lifetime
of SNe Ia, that is, $t_{\rm Ia}$ should be rather long, $\ga 1$ Gyr.

The dependence of the [Fe/H] distribution on $t_{\rm Ia}$ is not as
simple as the abundance ratio.  These figures show that the lifetime
$t_{\rm Ia}$ seems to determine the width of [Fe/H] distribution, that
is, a shorter $t_{\rm Ia}$ makes the [Fe/H] distribution narrower.  It
could be related to the epoch of star formation.  In the shorter $t_{\rm
Ia}$ case, many intermediate stars explode as SNe Ia and begin to
release metals into the interstellar medium before the major epoch of
star formation compared with the reference model, therefore most of the
stars have similar iron abundances, which makes the [Fe/H] distribution
narrow.  In contrast, in the longer $t_{\rm Ia}$ case, some of the stars
form in iron-poor environments before most of the iron is released, and
some of the stars form in iron-rich environments because at late epochs
the gas fraction is low and iron is released into gas-poor systems.
Thus the [Fe/H] distribution has a wider width.

\subsection{Star formation timescale}
\begin{figure}

\plotone{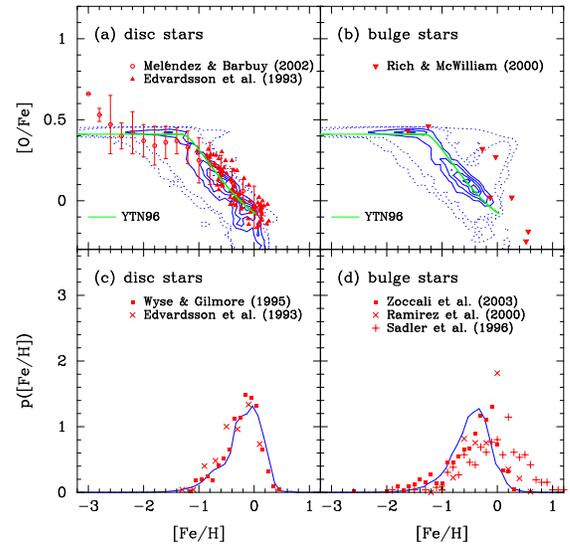}

\caption{The same as Figure \ref{fig:fid} but for $\tau_{*}^{0}=2.6$
 Gyr.}

\label{fig:sft1}
\end{figure}

\begin{figure}

\plotone{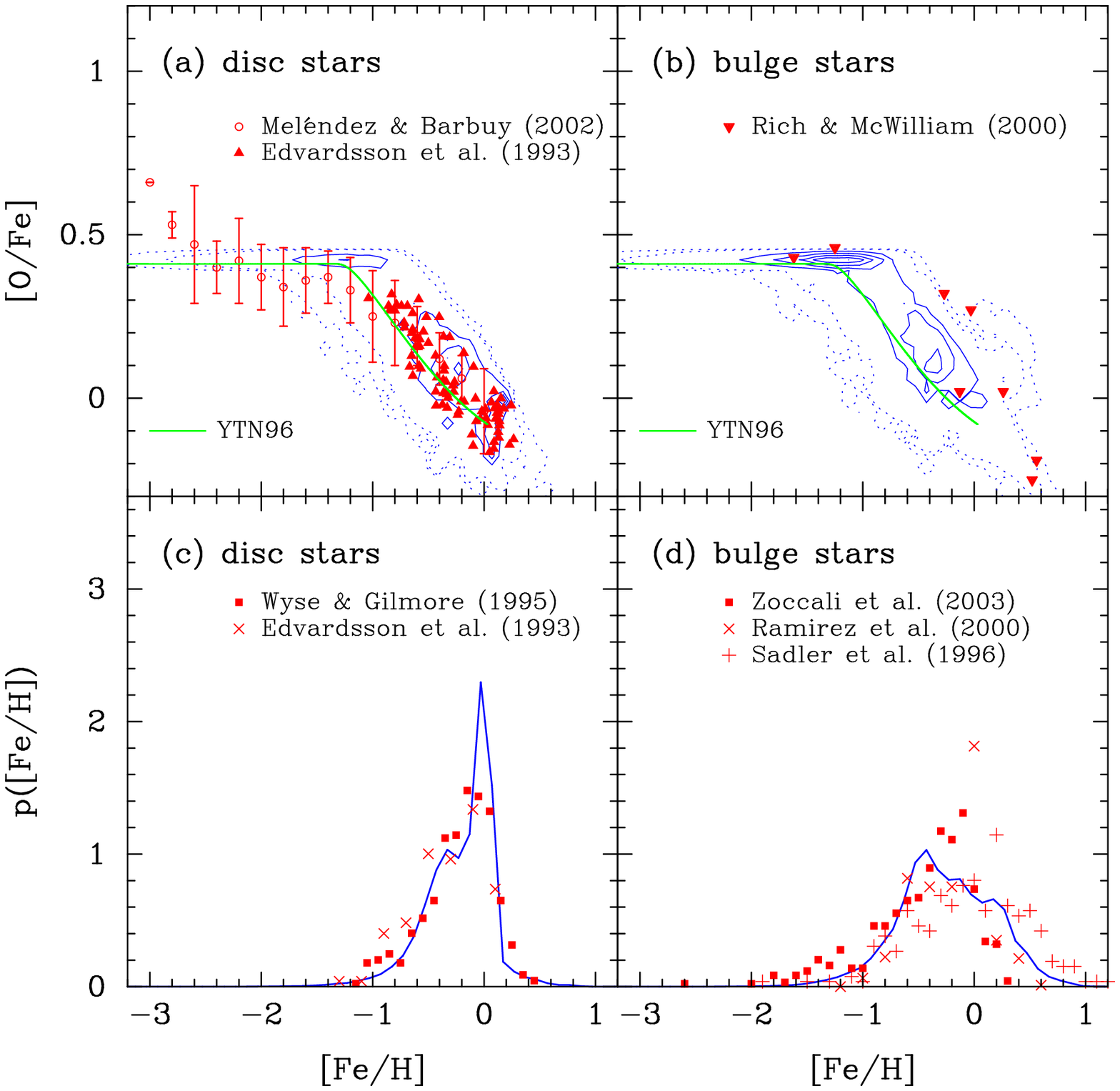}

\caption{The same as Figure \ref{fig:fid} but for $\tau_{*}^{0}=0.65$
 Gyr.}

\label{fig:sft2}
\end{figure}

Next, we investigate the dependence on star formation timescale.  In
Figures \ref{fig:sft1} and \ref{fig:sft2}, we show models which are the
same as the reference model, except for having twice and half the star
formation timescale of the reference model, $\tau_{*}^{0}=2.6$ and 0.65
Gyr, respectively.


\begin{figure}

\plotone{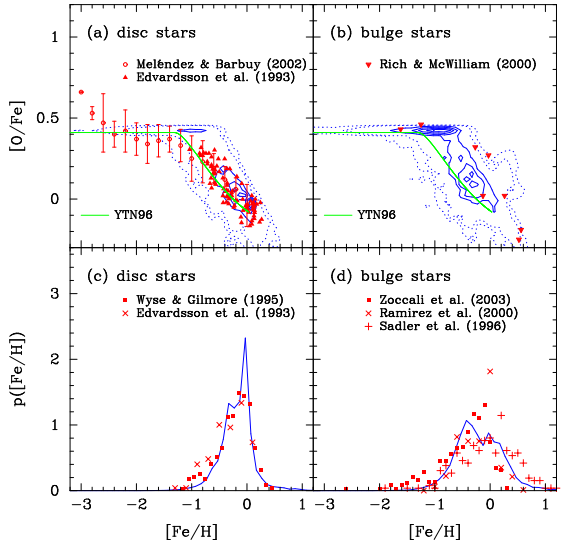}

\caption{The same as Figure \ref{fig:fid} but for the DSF model.}

\label{fig:sft3}
\end{figure}

\begin{figure}

\plotone{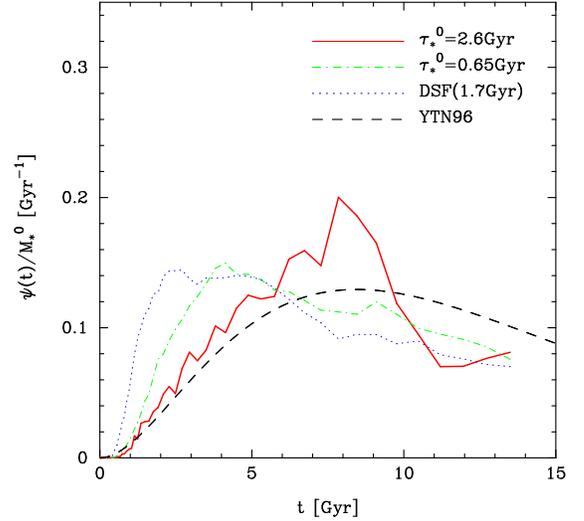}

\caption{Star formation histories.  The solid, dot-dashed and dotted
 lines indicate models with $\tau_{*}^{0}=2.6$Gyr and 0.65Gyr, and a
 model with DSF, respectively, corresponding to Figures \ref{fig:sft1},
 \ref{fig:sft2} and \ref{fig:sft3}.  The dashed line denote the infall
 model.  }

\label{fig:sfr2}
\end{figure}

As seen in Figure \ref{fig:sft1}b, in the longer star formation
timescale case, the break point in [O/Fe]--[Fe/H] plane for bulge stars
moves towards lower [Fe/H], because the star formation epoch shifts to
lower redshift and then the enrichment due to SNe Ia effectively begins
earlier.  In the case of a shorter star formation timescale, by
contrast, the break point moves towards higher [Fe/H].  Thus we can say
that the position of the break point is generally determined by a
combination of the lifetime of SNe Ia and the star formation timescale,
such as $t_{\rm Ia}/\tau_{*}^{0}$.  In the longer $\tau_{*}^{0}$ case,
there is a broader tail showing oxygen-enhancement at [Fe/H]$\sim 0$
than in the reference model.  This is for the same reason as the case of
shorter $t_{\rm Ia}$.  The difference from the case with shorter $t_{\rm
Ia}$ is that the distribution is not U-shaped but an inverse triangle,
because the SNe Ia continue enrichment at lower redshifts.  Disk stars
also have the same dependence on the star formation timescale, but we
can clearly see those tendencies for bulge stars.

The [Fe/H] distribution shows a different tendency from the dependence
on $t_{\rm Ia}$.  In the case of longer and shorter $\tau_{*}^{0}$, the
distributions typically { but slightly} have lower and higher [Fe/H],
respectively, { at least for bulge stars.}  Generally a longer star
formation timescale results in a larger cold gas fraction.  Then, with
the longer star formation timescale, the metallicity becomes lower than
in a model with a shorter star formation timescale.  In addition, such a
longer star formation timescale decreases the amount of stars, which
means that fewer metals are released from SNe Ia.  Thus the star
formation timescale has a different effect from the lifetime of SNe Ia.

In the reference model, the star formation timescale is assumed to be
constant with redshift.  It is interesting to see how the redshift
dependence affects the chemical enrichment.  In \citet{ny04}, another
formulation for the star formation timescale is considered; it is
proportional to the dynamical timescale of galactic disk, which is
called the dynamical star formation (DSF) timescale model in that paper.
Because we assume the disk rotation velocity to be almost the same as
the circular velocity of dark halos, which becomes larger toward higher
redshift, the dynamical timescale becomes shorter toward higher
redshift.  Figure \ref{fig:sft3} shows the results of adopting the DSF
model, in which the star formation timescale at present is
$\tau_{*}^{0}=1.7$ Gyr, so as to produce an agreement of cold gas
fraction in spiral galaxies with observation.  We find that the position
of the break point in [O/Fe]--[Fe/H] plane slightly moves toward higher
[Fe/H], and the slope of [O/Fe] against [Fe/H] seems to be slightly
steeper.  In the DSF model, since most of stars form at higher redshift
than in the reference model due to the shorter star formation timescale
at high redshift, enrichment of iron happens quickly.  Thus the slope is
steeper and the typical [Fe/H] is larger than in the reference model,
but this effect is very small.

In Figure \ref{fig:sfr2}, we show star formation histories of those
three models.  The solid, dot-dashed and dotted lines indicate models
with $\tau_{*}^{0}$=2.6Gyr and 0.65Gyr, and the DSF model, respectively.
The dashed line denotes the infall model given by \citet{ytn96}.  The
major star formation epochs of the models of $\tau_{*}^{0}=2.6$Gyr and
0.65Gyr are later and earlier than the reference model shown in Figure
\ref{fig:sfr1}.  Because a star formation timescale at high redshift in
the DSF model is much shorter than that in the reference model, the
major star formation epoch moves toward even earlier epoch than the
model of $\tau_{*}^{0}=0.65$Gyr.  From this figure, thus we can reach
the same conclusion as analyses by the infall model, that is, when major
star formation finishes before iron enrichment due to SNe Ia becomes
efficient, the break point in [O/Fe]--[Fe/H] moves toward higher [Fe/H]
because even SNe II alone sufficiently enrich stars.

As long as we stay varying only the lifetime of SNe Ia, $t_{\rm Ia}$,
and the star formation timescale, $\tau_{*}$, our SA model seems to
provide quite similar conclusions to the classical infall model.  This
apparent coincidence, however, breaks when the SN feedback is considered
as shown in the next subsection.

\subsection{Supernova feedback}
It has been widely realized that SN feedback significantly affects
galaxy properties.  Strong energy feedback suppresses chemical
enrichment due to SNe II because star formation is suppressed by
expelling cold gas, and also most of the metals are expelled into the
hot gas \citep{kc98, ng01}.  Thus it is worthwhile to investigate how SN
feedback affects the enrichment due to both SNe II and SNe Ia.

\begin{figure}

\plotone{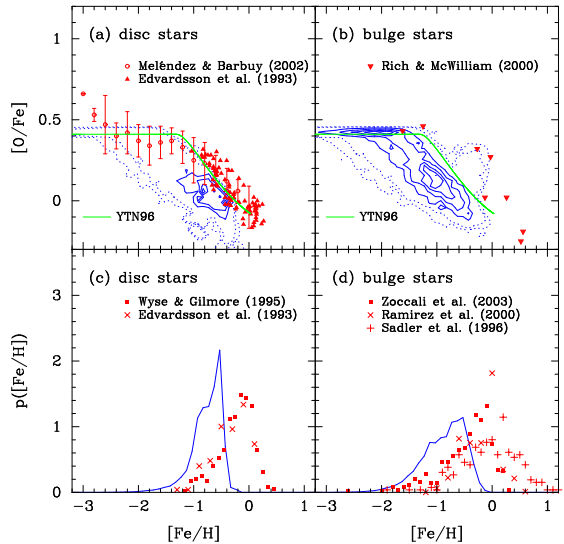}

\caption{The same as Figure \ref{fig:fid} but for $V_{\rm hot}=280$
 km~s$^{-1}$.}

\label{fig:snfb1}
\end{figure}

\begin{figure}

\plotone{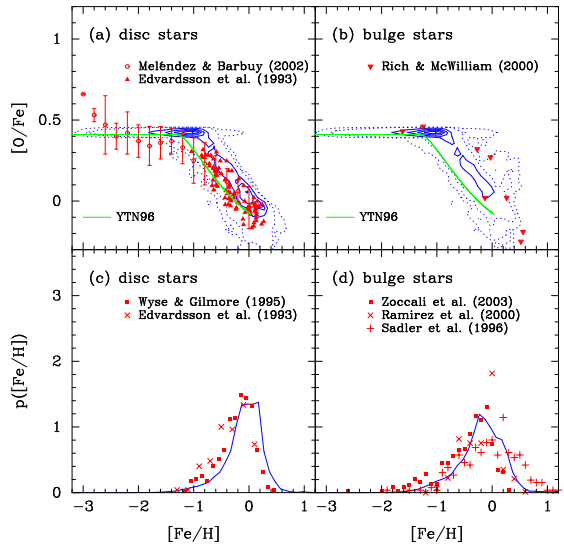}

\caption{The same as Figure \ref{fig:fid} but for $V_{\rm hot}=70$
 km~s$^{-1}$.}

\label{fig:snfb2}
\end{figure}

In Figures \ref{fig:snfb1} and \ref{fig:snfb2}, we show the same figures
as Figure \ref{fig:fid} but for $V_{\rm hot}=280$ (strong feedback) and
70 (weak feedback) km~s$^{-1}$, respectively, where $V_{\rm hot}$
determines the typical scale at which SN feedback becomes effective,
{ that is, SN feedback strongly affects galaxies with $V_{\rm
disk}<V_{\rm hot}$.}  As clearly shown in the bottom panels, the [Fe/H]
distributions for both disk and bulge stars are strongly affected by SN
feedback.  Strong feedback greatly suppresses formation of metal-rich
stars and thus the chemical enrichment.  In contrast, there are too few
metal-poor stars in the case of weak SN feedback compared with the
observed distributions, while the difference from the reference model is
smaller than that of the strong SN feedback case.  It should be noted
that our MW-like galaxies have $V_{\rm disk}\simeq 220$ km~s$^{-1}$ at
low redshift.  The values of $V_{\rm hot}$ for both the reference model
and the weak SN feedback case are less than the rotation velocity, while
in the strong SN feedback case $V_{\rm hot}$ is larger than the rotation
velocity.  That means that SN feedback is very efficient even at low
redshift in the strong feedback case.  Therefore the difference of the
strong SN feedback case from the reference model is larger.

The stronger the SN feedback is, the lower the [Fe/H] at the break point
is.  Consequently, strong and weak SN feedback seem to have similar
effects in the [O/Fe]--[Fe/H] plane to those of short and long $t_{\rm
Ia}$, and long and short $\tau_{*}^{0}$.  Since strong SN feedback
suppresses the chemical enrichment due to SNe II, the enrichment due to
SNe Ia begins at a lower [Fe/H].  These apparently correspond to longer
and shorter star formation timescales, respectively.

\begin{figure}

\plotone{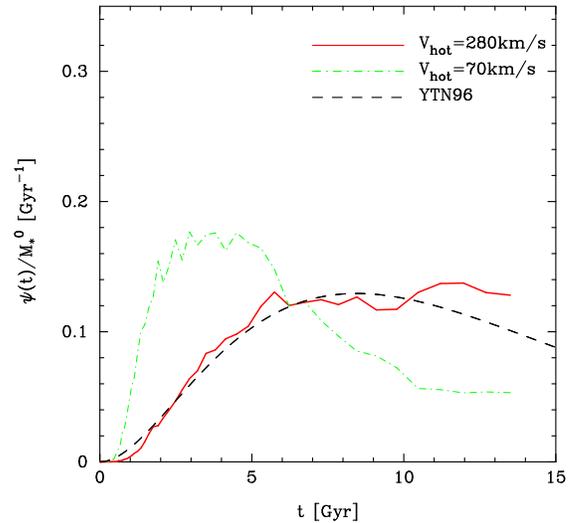}

\caption{Star formation histories.  The solid and dot-dashed lines
 indicate models with $V_{\rm hot}=280$ and 70 km~s$^{-1}$,
 respectively.  The dashed line denotes the infall model.  }

\label{fig:sfr3}
\end{figure}

Figure \ref{fig:sfr3} shows star formation histories of the strong
($V_{\rm hot}=280$km~s$^{-1}$) and weak ($V_{\rm hot}=70$km~s$^{-1}$)
feedback models indicated by the solid and dot-dashed lines,
respectively.  The dashed line denotes the infall model.  The strong
feedback model shows a similar star formation history to that of the
long star formation timescale model (the solid line in Figure
\ref{fig:sfr2}).  Both models have break points in [O/Fe]--[Fe/H] plane
at slightly lower [Fe/H] than that in the reference model.  The peak of
the [Fe/H] distribution of the strong feedback model, however, moves
toward lower [Fe/H] compared with the long star formation timescale
model.

The weak feedback model shows a similar star formation history to that
of the DSF model (the dotted line in Figure \ref{fig:sfr2}).  The break
points of both model move toward high [Fe/H] compared with the reference
model.  Again, the shift to higher [Fe/H] in the [Fe/H] distribution of
the weak feedback model is slightly enhanced compared with the DSF
model.

This is because SN feedback strongly affects chemical enrichment.  From
equations (\ref{eqn:Mcold})-(\ref{eqn:Zhot}), we obtain a mass-weighted mean
stellar metallicity,
\begin{equation}
 \langle Z_{*}(t)\rangle=Z_{c}^{0}+\frac{\alpha y}{\alpha+\beta}\frac{1-e^{-x}-xe^{-x}}{1-e^{-x}},
\end{equation}
where $x=(\alpha+\beta)t/\tau_{*}$ and $Z_{c}^{0}$ is an initial
metallicity of gas.  At the limit $t/\tau_{*}\to\infty$, the mean
metallicity becomes
\begin{equation}
 \langle Z_{*}(\infty)\rangle\to Z_{c}^{0}+\frac{\alpha y}{\alpha+\beta}.\label{eqn:Zbeta}
\end{equation}
This shows that a strong SN feedback ($\beta\gg 1$) suppresses chemical
enrichment due to SNe II.  { The effective chemical yield taking into
account SN feedback can be written as $y_{\rm eff}=y/(1+\beta/\alpha)$.}
The effect of SN feedback on chemical enrichment is the reason why, for
example, the strong feedback model and the long star formation timescale
model predict different [Fe/H] distributions in spite of their similar
star formation histories.  Therefore, similar star formation histories
do not always provide similar chemical enrichment histories.  Gas
outflow induced by SNe II plays a critical role in galaxy evolution.

\section{SUMMARY}
We have explored chemical enrichment due to both SNe II and SNe Ia in
Milky Way-like galaxies in the semi-analytic galaxy formation model.
Our treatment of SNe Ia is fully consistent with the galaxy formation
model, that is, we solve for the recycling of each element among stars,
cold gas and hot gas based on a $\Lambda$CDM model.  It is important to
follow the chemical enrichment in the framework of the hierarchical
galaxy formation because there are many non-trivial effects on the
chemical enrichment caused not only by mergers of galaxies but by
varying efficiencies of the SN feedback dependent on the depth of
gravitational potential wells of sub-galactic clumps.  As a first
attempt at constructing such a consistent model, we have assumed that
all SNe Ia have the same lifetime, $t_{\rm Ia}$, that abundance patterns
of metals from SNe Ia and SNe II are always the same, and that massive
stars instantaneously explode as SNe II and release metals.  This is a
natural extension of the work by \citet{ytn96}, in which they
self-consistently modeled the chemical enrichment due to SNe Ia and SNe
II.

We have picked out galaxies in dark halos with $V_{\rm circ}=220$
km~s$^{-1}$ and having a similar luminosity to the MW.  We have found
that when we impose $t_{\rm Ia}$=1.5 Gyr, the predictions of our model
for such MW-like galaxies agree well with observations of the chemical
composition of solar-neighborhood stars both in the stellar distribution
in the [O/Fe]-[Fe/H] plane and in the { iron MDF}.  We would like to
stress that the other model parameters such as the star formation
timescale and SN feedback are the same as in the fiducial model of
\citet{ny04}, in which they found that the model reproduce well many
aspects of observed galaxies, such as luminosity functions, cold gas
fractions and sizes of galaxies for local galaxies, and the surface
brightnesses, velocity dispersions, mass-to-light ratios and
metallicities of local dwarf spheroidals.  This work also shows that the
classical G-dwarf problem \citep{v62, s63, pp75} is fully resolved in
the framework of hierarchical formation of galaxies, in which the infall
term introduced in the infall models to avoid the G-dwarf problem is
naturally explained by the mixture of such physical processes as
clustering of dark halos, gas cooling and SN feedback.  Our model passes
the new tests, that is, { the iron MDF} and the abundance pattern of
metals, taking into account chemical enrichment due to SNe Ia.  As shown
in Section \ref{sec:depend}, the abundance pattern provides an
independent constraint on galaxy formation.  Therefore our results would
support the scenario of hierarchical formation of galaxies.  While
observational data used in this work have been limited to
solar-neighborhood and bulge stars in the Milky Way, future observations
of stars in other galaxies will provide statistically better constraints
on galaxy formation.

In our model, disk and bulge stars are treated separately.  While the
observations of bulge stars still have uncertainties, our model also
reproduces well both the distribution of stars in the [O/Fe]--[Fe/H]
plane and the [Fe/H] distribution.  In particular, although there are
only a few bulge stars whose oxygen abundance is observed, our model
predicts oxygen-enhanced stars at [Fe/H]$\sim 0$, but they are not a
dominant fraction.  Increasing observational data will provide a strong
constraint on bulge formation.

As a by-product, we have obtained the luminosity function of galaxies in
Local Group-halos.  Recent high-resolution $N$-body simulations have
predicted many more dwarf scale dark halos than observed satellite
galaxies, and this has been considered to be a serious problem
\citep{kkvp99, m99}.  Such an overabundance problem has been solved in
the framework of the SA models taking into account effects of
reionization and incompleteness due to significantly low surface
brightness \citep{s02, bflbc02b}.  Our model also shows an ability to
solve this problem in a similar manner to those works.

The chemical yields we used are the same as those in \citet{ytn96}, in
which an infall model of monolithic collapse was used \citep{ayt92}.
The similarity between our hierarchical model and the monolithic cloud
collapse model suggests that spiral galaxies in a hierarchical universe
should have a similar formation history to that modeled by the
monolithic cloud collapse model, as shown in \citet{bcf96}.  Star
formation histories we computed also suggests the similarities.  It
does, however, not always mean that similar star formation histories
provide similar chemical enrichment histories.  As shown in Section 6.3
and more directly by equation (\ref{eqn:Zbeta}), SN feedback plays a
critical role in chemical enrichment.

In this paper we have concentrated on investigating the [O/Fe] relation
and the [Fe/H] distribution.  Since recent analyses of
solar-neighborhood stars have revealed that the age-metallicity
relation has a large scatter, it will be useful as a next step to see
whether our model can reproduce this relation as well as the scatter in
it \citep{ia02}.  Furthermore, while the analysis in this paper focused
only on MW-like galaxies, our SA model has the potential ability to
investigate other systems simultaneously.  To investigate abundance
ratios in such systems as the intracluster medium, the stars composing
elliptical galaxies, and damped Ly-$\alpha$ systems will provide
independent, important clues to understanding galaxy formation.  The
cosmic explosion rate of SNe Ia will also give a new insight into both
galaxy formation and observational cosmology.  We will pursue these
topics in future papers.

\section*{ACKNOWLEDGMENTS}    
We would like to thank Takuji Tsujimoto for useful suggestion and Cedric
G. Lacey for reading our paper carefully.  We also acknowledge support
from the PPARC rolling grant for extragalactic astronomy and cosmology
at Durham and from the Japan Society for the Promotion of Science for
Young Scientists (No.00207 and 01891).

\end{document}